\documentclass[aps,amsfonts,nofootinbib,twocolumn,showpacs]{revtex4}
\usepackage{amssymb}
\usepackage{graphicx}
\usepackage{amsmath}
\def\be{\begin{equation}}
\def\ee{\end{equation}}
\def\bea{\begin{eqnarray}}
\def\eea{\end{eqnarray}}

\def\om{\Omega_{\text{m}}}

\def\gta{ \lower .75ex \hbox{$\sim$} \llap{\raise .27ex \hbox{$>$}} }
\def\lta{ \lower .75ex\hbox{$\sim$} \llap{\raise .27ex \hbox{$<$}} }
\def \Ol {$\Omega_\Lambda$}
\def \Om {$\Omega_{\rm M}$}
\def\chs{\mbox{$\chi^2$}}

\newcommand{\kmsmpc}{{\rm \, km\, s}^{-1}{\rm Mpc}^{-1}}
\newcommand{\beq}{\begin{equation}}
\newcommand{\eeq}{\end{equation}}
\newcommand{\beqa}{\begin{eqnarray}}
\newcommand{\eeqa}{\end{eqnarray}}
\newcommand{\ow}{\Omega_w}

\def\fun#1#2{\lower3.6pt\vbox{\baselineskip0pt\lineskip.9pt
  \ialign{$\mathsurround=0pt#1\hfil##\hfil$\crcr#2\crcr\sim\crcr}}}
\usepackage{txfonts}
\usepackage{graphicx}

\begin{document}
\title{A Cosmic Model Parameterizing the late Universe}
\author{Xin-He Meng$^{1,4}$}
\email{xhm@nankai.edu.cn}
\author{Meng Su$^{2,3}$}
\email{mengsu@cfa.harvard.edu}
\author{Zheng Wang$^{5}$}
\affiliation{$^1$Department of physics, Nankai University, Tianjin
300071, P. R. China} \affiliation{$^2$Astronomy Department, Harvard
University, 60 Garden Street, Cambridge, MA 02138, USA}
\affiliation{$^3$Department of Astronomy, School of Physics, Peking
University, Beijing 100871, P. R. China}\affiliation{$^4$BK21
Division of Advanced Research and Education in physics, Hanyang
University, Seoul 133-791, Korea and Department of physics, Hanyang
University, Seoul 133-791, Korea}\affiliation{$^5$National
Laboratory for Information Science and Technology, Tsinghua
University, Beijing, 100084, P. R. China}
\date{\today}
\begin{abstract}
A simple speed-up cosmology model is proposed to account for the
dark energy puzzle. We condense contributions from dark energy and
curvature term into one effective parameter in order to reduce
parameter degeneracies and to find any deviation from flat
concordance $\Lambda$CDM model, by considering that the
discrimination between dynamical and non-dynamical sources of cosmic
acceleration as the best starting point for analyzing dark energy
data sets both at present and in future. We also combine recent Type
Ia Supernova (SNIa), Cosmic Microwave Background (CMB) and Baryon
Oscillation (BAO) to constrain model parameter space. Degeneracies
between model parameters are discussed by using both degeneracy
diagram and data analysis including high redshift information from
Gamma Ray Bursts (GRBs) sample. The analysis results show that our
model is consistent with cosmological observations. We try to
distinct the curvature effects from the specially scaling dark
energy component as parameterized. We study the linear growth of
large scale structure, and finally show the effective dark energy
equation of state in our model and how the matter component
coincidences with the dark energy numerically.
\end{abstract}


\maketitle

\section{Introduction}
It is now well-established that the expansion of our universe is
currently in an accelerating phase, supported by the most direct and
robust evidence from the redshift - apparent magnitude measurements
of the "cosmic lighthouse" type Ia supernova \cite{Perlmutter}, and
indirect others such as the observations of Cosmic Microwave
Background (CMB) by the WMAP satellite
\cite{Spergel,jarosik,hinshaw,page,boomerang,cbi,kuo,sa,planck}, and
large-scale galaxy surveys by 2dF and SDSS
\cite{cole,sdss1,sdss2,tegmark}. Under the assumption that general
relativity is valid on cosmological scale, the combined analysis of
different observation data sets indicates a spatially-flat universe
with about 70\% of the total energy content of the universe today as
so called dark energy with effectively negative pressure responsible
for the accelerating expansion (see Ref.~\cite{Peebles} for reviews
on this topic). Among multitudinous candidates of dark energy
models, the "simplest" and theoretically attractive one might be the
so called vacuum energy, i.e. $\rho_{\Lambda} = \Lambda/8 \pi G$,
where $\Lambda$ is the cosmological constant, which has been long
considered as a leading candidate and works quite well on explaining
observations through out the history of our universe at different
scales. But the origin or mechanisms responsible for the cosmic
accelerating expansion are not very clear. On the other hand, some
authors suggest that maybe there does not exist such mysterious dark
component, but instead the observed cosmic acceleration is a signal
of our first real lack of understanding of gravitational physics
\cite{Lue} on cosmic scale. An example is the braneworld theory with
the extra dimensions compactified or non-compactified
\cite{Dvali,rs,hw,ms2}. Consequently, finding the different
cosmological implications to distinguish modified gravity models and
dark energy scenario from observations is essentially fundamental to
physically understanding of our universe\cite{distinguish}.

Along with the matter (mainly cold dark matter) component and
possible curvature term, the mysterious dark energy dominates the
fate of our universe (we do not consider the radiation component
contribution as it is supposed to be very tiny for the current
universe evolution, at least for the present discussion interests).
Ironically so far we do not know much to either of them, even full
of puzzling to some extends. So any progress or reasonable
understanding to each of them is undoubtedly valuable. Specifically,
the quest to distinguish between dark energy and modified gravity
scenario and further to differentiate cosmological constant and
dynamical dark energy models from observations has become the focus
of cosmology study since it holds the key to new fundamental
physics.

Although we have built up a successful parametrization to describe
the properties and evolution of our universe, and in principle
distinct dark energy models live at different sub-space of fully
descriptive multi-dimension parameter space, due to serious
degeneracies among different parameters, we cannot get tight enough
constraints from observations by global fitting various
observational data sets. One way to extract useful information from
observation data and get hints for fundamental physics from
cosmology study is to reduce the dimension of parameter space (thus
reduce the parameter degeneracies) with particular purpose in mind
without apparently biased input to model parametrization. Since we
have not found any evidence of inconsistency of standard
$\Lambda$CDM model, including more parameters which describe
detailed properties of each component if the "cosmic pie" will
complicate the situation to constrain model parameters.

In order to find deviation of dark energy equation of state
parameter $w$ from $-1$ (e.g. evidence of non-cosmological constant
dark energy), the assumption of a flat universe is widely accepted
in the literature with claims that curvature is negligible from
inflation predictions and with emphasis on combined analysis results
\emph{with} prior assumption $w=-1$. On the other hand, typically
one looks for evidence of dynamical dark energy in the absence of
spatial curvature to get better constraints (for an exception, see
\cite{curvature1}). It has been concluded in \cite{curvature} that
the non-curvature assumption can induce critically large errors in
reconstructing the dark energy equation of state even if the true
cosmic curvature is on sub-percent level. These claims motivate us
proposing a parameterized dark component term to mimic the effective
contributions from either dark energy or curvature term plus the
dark energy (It is also possible that the parameterized term we
postulate may be from a fundamental theory or reasonably modified
gravity model we are seeking), besides the conventional matter term.

In the first step, it is reasonable to introduce only one parameter
which stands for any kind of deviation from standard cosmology
model. In some limit case, it should be reduced to the simple four
dimensional (4D) $\Lambda$CDM cosmology. The constraint on this
parameter from observations should provide insightful hints to
further explore fundamental physics.

In the next section, we propose a simple cosmic parametrization for
the current universe, a parameterized model for the later evolution of our universe. In section 3 we give various cosmic probes to
this model, with comparison to the DGP model Universe\cite{Dvali} and
the concordance model with a cosmological constant, i.e, the
$\Lambda$CDM model, with the hope to locate new features to this new
model. Then in section 4, we discuss the new degeneracies between
the parameters we introduced and dark matter content. The possible
constraints from high redshift observations are also discussed. The
last section devotes discussions and conclusions for the general
framework studies to this present model.

\section{a parameterized late Universe}

Firstly we summarize some basics in Standard cosmology Model (ScM)
 as a preparation for our present work. The ScM starts with a solution of Einstein's equation in the 4D
Friedmann-Robertson-Walker (FRW) spacetime with cosmo assumed full
of perfect fluid in large cosmic scale, mainly described by the
Friedmann equation for the Hubble parameter evolution (expansion
rate of the Universe) as
\begin{equation}
H^2=(\dot a/a)^2=\rho/3,\label{eq:rab}
\end{equation}
where the global scale factor a(t) describes the cosmic evolution
history and the isotropic density $\rho$ satisfies the fluid
continuum (conserved) equation. A complete expression for the
Hubble expansion rate, which extends the FRW solutions to include
all cosmic components so far we know is given by (and we take
conventions hereafter, that is we work in natural units where
$c=8\pi G=1$ and $a_0=1$),

\begin{eqnarray}
H^2 &=& Ma^{-3}-ka^{-2}+Ra^{-4}+\frac{\Lambda}{3}\nonumber\\
&=&
H_0^2[\Omega_ma^{-3}-\Omega_ka^{-2}+\Omega_Ra^{-4}+\Omega_\Lambda],
\end{eqnarray}
where the subscript 0 indicating today's value, curvature fraction
$\Omega_k=k/H_0^2$, similarly to matter component fraction
$\Omega_m$, cosmological constant (Dark Energy) contribution
$\Omega_\Lambda$ and the radiation part $\Omega_R=R/H_0^2$ can be
negligible today when compared to the mainly dark components.

In the year 2000, G. Dvali, G. Gabadadze and M. Porrati proposed a
new model that can mimic the 4D Newton potential (with same scaling)
in short ranges while it makes a 5D gravity model (the corresponding
potential scaling differently from the conventional 4D Newton
potential) in long ranges. It turns out interesting to compare this
ScM to the DGP model, an extended ScM that in the simplest flat
geometry case an additional term $H_c$ contributes a cosmic scale
related effect that deviates the common framework at large distances
and we hope to know when it functions. The Friedmann equation for
Hubble expansion in the DGP model (we take the self-accelerating
branch solution with the plus sign in front of the root term) reads
as
\begin{equation}
H^2-k/a(t)+(H^2-k/a(t))^{1/2}/r_c=\rho/3
\end{equation}
where the Hubble parameter or expansion rate $H^2=(\dot a/a)^2$. In the DGP model, gravity is trapped on a four-dimensional (4D) brane world at short
distances, but is able to propagate into a higher-dimensional space at large distances. For the convenient comparasions we take its flat geometrical form
\begin{equation}
H^2=(\dot a/a)^2=\rho/3+{H_c}^2\label{eq:rab}
\end{equation}
where the effective term ${H_c}^2=H/r_c$ that we treat as a parameter to be fitted in
this work and the cross-over length scale defined by Planck mass
over a 5D scale $r_c=M_{Pl}^2/{2M_5^3}$.

Conventionally, the redshift is defined by $z=1/a-1$, thus
$a^{-1}=1+z$. Compared with the expression of $H^2$ for the
power-law $\Lambda$CDM model, we have known that from this 4D
cosmological model with cosmological constant, the cosmological
observation data analysis can be nicely accomodated/explained with
curvature contribution near zero, so named as the concordant model.
While the global data fitting successful we are still left with the
curiosity that whether the cosmic curvature term is really zero or
it can be effectively described by the accumulated effects from the
comsic un-known dark components\cite{aa}?

\begin{table}[h]
\caption{\label{tab:t1} Physics meanings in the Friedmann
evolution Eq.}
\begin{ruledtabular}
\begin{tabular}{ccc}
Functions or constants & Physical meanings & Terms in $H^2$\\
\hline
$\dot{a}$ & "Expansion velocity" &\\
$M$ & Matter (dust) & $\Omega_m(1+z)^3$\\
$k$ & Curvature & $\Omega_k(1+z)^2$\\
$R$ & Radiation & $\Omega_R(1+z)^4$\\
$\Lambda$ & Cosmological constant & $\Omega_\Lambda$
\end{tabular}
\end{ruledtabular}
\end{table}

The reduction to the $\Lambda$CDM model can be also realized in a
more economic form as parameterized below
\begin{equation}
H(z)^2=H_0[(\Omega_m)a^{-3}+(1-\Omega_m)a^{B-2}].
\end{equation}
where B is a parameter to be determined by data fittings and
obviously $B=2$ corresponds to the $\Lambda$CDM model, we call the
term including B parameter effective dark energy (EDE) in the
following. We can compare it with the $\Lambda$CDM model in the flat
geometry where the Hubble parameter $\tilde{H}(z)$ is
\begin{equation}
\tilde{H}(z)=H_0[\Omega_m(1+z)^3+1-\Omega_m].
\end{equation}
Thus, the general case with all possible components we understand
so far reads as
\begin{equation}
H(z)^2=\tilde{H}(z)^2-k(1+z)^2+R(1+z)^4.
\end{equation}
Of course we can encode relevant physics in the parameter $B$, but
in the 4D Universe with cosmological constant, each arbitrary
parameter and term separately possesses concrete physics meanings
compared with the 5D DGP model. So we employ various cosmological
tests to see what physics the parameter B may stand for, the
effective effects from both curvature and dark energy, or curvature
term only with $B=0$ or dark energy alone in the flat spacetime
geometry. Finally, we want to ask how well can we distinguish the
curvature effects from the dark energy component?

Among dark energy candidate models, among which the modified gravity
or decaying cosmology term models can effectively describe possibly
dark matter interacting with dark energy\cite{we, dem}, to which we
also expect this parameterized model can help. The detailed
discussion on this topic is beyond the scope of this paper.

We also note that there is a long standing issue on breaking
degeneracies between curvature and dynamical dark energy model
parameters. For example, CMB lensing information can effectively
help to break such
degeneracy\cite{lensing1,lensing2,lensing3,lensing4,lensing5} with
only already planned ground-based CMB polarization power spectrum
measurements. But the results depend on two strong assumptions, one
is that the ground-based CMB survey will be able to remove
foregrounds and systematics at a level sufficient to enable few
percent level measurements of the lensing B-mode polarization power,
another one is that the neutrino masses are fixed by oscillation
measurements and a theoretical assumption about the neutrino mass
hierarchy\cite{lensing5}. So even with ideal future measurements on
CMB lensing, our new parametrization still has its advantages.

\section{Observational Constraints}

In this section, we study the cosmological constraints on our model
parameter spaces. There are several methods which have been used or
proposed to constrain cosmological parameters in the literature,
e.g. Type Ia supernove, CMB, linear power spectrum and higher order
statistics of large scale structure\cite{bispectrum}, Lyman-alpha
forest\cite{LF}, Alcock-Paczy\'{n}ski (AP) effect\cite{su3},
weak$/$strong gravitational lensing\cite{WL}, Gamma ray
bursts$/$ultra-compact radio sources as standard
candles$/$rulers\cite{su1,su2,Schaefer,radiosources}, X-ray cluster
baryon fraction versus redshift test\cite{fgas}, Hubble parameter
measurements on different redshift\cite{hubbleparameter}, cluster
counting\cite{clustercount} and so on. In principle, in order to get
self-consistent parameter constraints, one should do a global
fitting on whole cosmological parameter space with properly chosen
observational data sets. However, global fitting is time$/$CPU
consuming and it is hard to analyze the degeneracies on parameter
spaces.

In this paper, in the first setup to look at our model parameter
space and to analyze the parameter degeneracies clearly, we use
recent SNe Ia gold sample \cite{Riess} and SNLS data\cite{SNLS}, and
combine with information from WMAP three year data and SDSS analysis
results in our explorations. The SNLS sample consists of 44 nearby
(0.015$<$z$<$0.125) objects assembled from the literature, and 73
distant SNIa (0.15$<$z$<$1.00) discovered and carefully followed
during the first year of SNLS group\cite{SNLS}. For the cosmological
fits, two of the SNLS data points were excluded because they are
outliers in the Hubble diagram. For the SNIa data, the distance
modulus is defined as
\begin{equation}
M -m = 5\log d_L + 25.
\end{equation}
Here $d_L$ is the luminosity distance in units of Mpc which is
written as
\begin{equation}
d_L = \frac{1+z}{\sqrt{|\Omega_k|} } \mathcal{S}  \left(
\sqrt{|\Omega_k|} \int_0^z \frac{dz'}{H(z') /H_0} \right)
\end{equation}
where $\mathcal{S}$ is defined as $\mathcal{S}(x) = \sin (x)$ for a
closed universe, $\mathcal{S}(x) = \sinh (x)$ for an open universe
and simply $\mathcal{S}(x) = x$ with non-curvature universe.

To further break the parameter degeneracies, it is useful to study
the combined constraints with other cosmological observations, we
make use of the CMB shift parameter which includes the whole shift
information of CMB angular power spectrum. It is defined as
\begin{equation}
R  = \frac{\Omega_m}{\sqrt{|\Omega_k|} } \mathcal{S}  \left(
\sqrt{|\Omega_k|} \int_0^{z_l} \frac{dz'}{H(z')/H_0} \right)
\end{equation}
where $z_l=1089$, the redshift of the epoch of the recombination.
The shift parameter is constrained to be $R = 1.70 \pm 0.03$ from
the three-year WMAP result, CBI and ACBAR\cite{R}. The CMB shift
parameter contains the main information for the scale of the first
acoustic peak in the TT spectrum, and is the most relevant one for
constraining dark energy properties as it is not sensitive to
different dark energy models. Since we only consider the shift
parameter which is determined only by the background evolution for
the constraint from CMB, we do not need to include the effect of the
fluctuation of dark energy. In this paper, by using shift parameter,
we can confine ourselves to considering the effects of the
modification of the background evolution alone.

And we also use the information from observation of baryon
oscillation acoustic peak which has been detected from the SDSS
luminous red galaxy sample\cite{sdss2}. The quantity we use to
constrain the cosmological parameters in this paper is defined as
\begin{equation}
A = \frac{\sqrt{\Omega_m}}{(H(z_1)/H_0)^{1/3}} \left[ \frac{1}{z_1
\sqrt{|\Omega_k|} } \mathcal{S}  \left( \sqrt{|\Omega_k|}
\int_0^{z_1} \frac{dz'}{H(z')/H_0} \right) \right]^{2/3}
\end{equation}
where $z_1=0.35$ and $A$ is measured as $A=0.469\pm 0.017$
\cite{sdss2}. Recently, the new SDSS LRG data were released, and the
corresponding power spectrum was analyzed. The BAO peaks are clearly
seen in the power spectrum, which, together with the overall shape,
put tight constraints on model parameters\cite{sdss3}.

For the fitting methodology, we use the standard \chs\ minimization
method. It is well known that parameter estimates depend sensitively
on the assumed priors on other parameters. In our study, we choose
the allowed range of the Hubble constant $H_0=72\pm8\enspace\kmsmpc$
resulting from the Hubble Space Telescope Key Project with a uniform
prior\cite{Hubble}, and marginalize over $H_0$ to get
two-dimensional constraints for our parameter space.

\begin{figure}[tbp]
\centering
\includegraphics[scale=0.8] {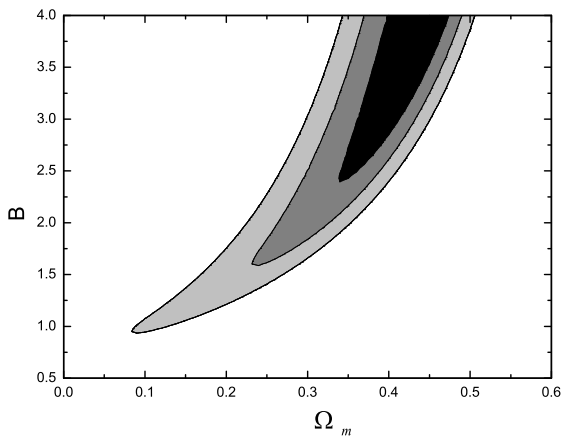}\quad
\caption{The black, grey, and light grey region shows the 1, 2, and
3 $\sigma$ confidence level contours of $\Omega_m-B$ parameter space
respectively on using the SNIa Gold data from Riess et al. }
\label{fig:Figure1}
\end{figure}

Fig.1 shows confidence-level contours on $\Omega_m-B$ parameter
space using the SNIa Gold sample. The black, grey, and light grey
region shows the 1, 2, and 3 $\sigma$ confidence level contours of
$\Omega_m-B$ parameter space respectively with the minimum
$\chi^2=158.42$ occurring at $\Omega_m=0.46$, and $B=4.62$. We note
that the best fit point is far from standard concordance cosmology,
but the 2 $\sigma$ confident contour is consistent with standard
concordance cosmology. The parameter degeneracy properties between
model parameters determine the configurations of constraint
contours. We will analyze the degeneracy properties on $\Omega_m-B$
parameter space in Fig.7. We note that the constraint results depend
sensitively on the prior assumptions that one adopts. A strong prior
can result in an overestimate on the power of a cosmological probe
or make a incorrect constraints on key parameters, bias our
judgement on model selection, thus improperly ruling out models.
Especially, it is also noted that factitious priors on $H_0$ can
result in strongly biased constraints.

\begin{figure}[tbp]
\centering
\includegraphics[scale=0.8] {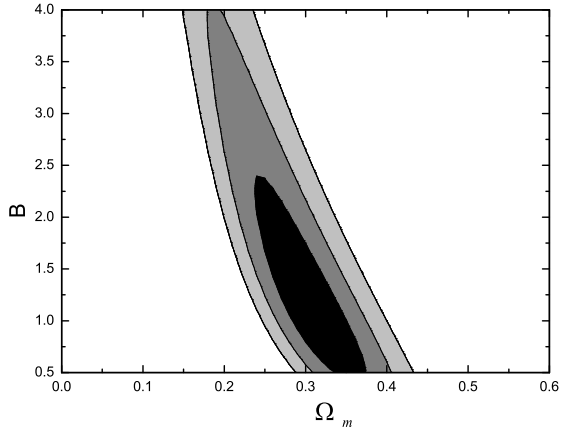}\quad
\includegraphics[scale=0.8] {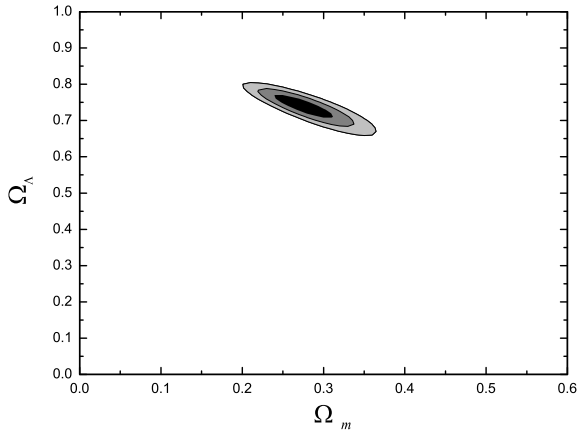}\quad
\includegraphics[scale=0.8] {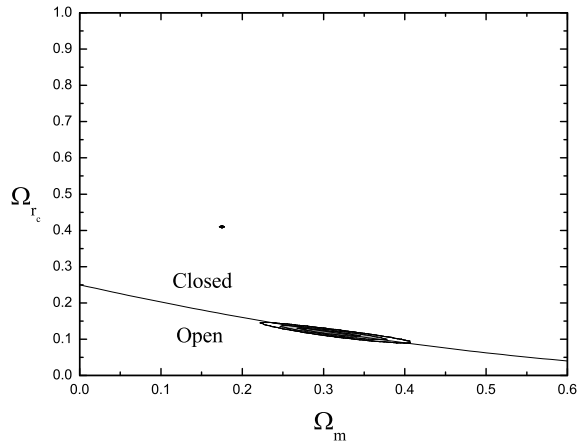}
\caption{The black, grey, and light grey regions in top, middle and
bottom figures show the 1, 2, and 3 $\sigma$ confidence level
contours of $\Omega_m-B$, $\Lambda$CDM and DGP model parameter
spaces, respectively, on combining CMB shift parameter from WMAP
three years data and BAO from SDSS} \label{fig:Figure2}
\end{figure}

In Fig.2, we show the constraint results from combining CMB shift
parameter with BAO from large scale structure of galaxies on
different cosmological models. The black, grey, and light grey
regions show the 1, 2, and 3 $\sigma$ confidence level contours of
EDE, $\Lambda$CDM and DGP model parameter spaces, respectively, with
the minimum $\chi^2$ occurring at $\Omega_m$=0.31, and $B$=1.10 for
EDE model. It has been clearly shown in the figure that even combing
information from CMB and BAO, which gives tight constraints on both
\Om -\Ol~ parameter space in $\Lambda$CDM model (middle sub-figure)
and \Om -$\Omega_{r_c}$(bottom sub-figure) parameter space in DGP
model, cannot constrain our model parameter space tightly. The
reason is that in $\Omega_m-B$ parameter space, CMB shift parameter
and BAO factor show similar degeneracy properties and thus cannot
break the 'banana' shape of constraint contours. Fortunately, we
find that the constraint contours on $\Omega_m-B$ parameter space by
using SNIa are almost perpendicular to contours from CMB+BAO
constraints as two sets of 'mirror bananas'(see figure 3 for
combined results). That means that in our model luminosity-distance
measurements from SNIa contributes considerably to the cosmological
constraints comparing with $\Lambda$CDM model and DGP model due to
different degeneracy properties shown on each parameter space.

\begin{figure}[tbp]
\centering
\includegraphics[scale=0.8] {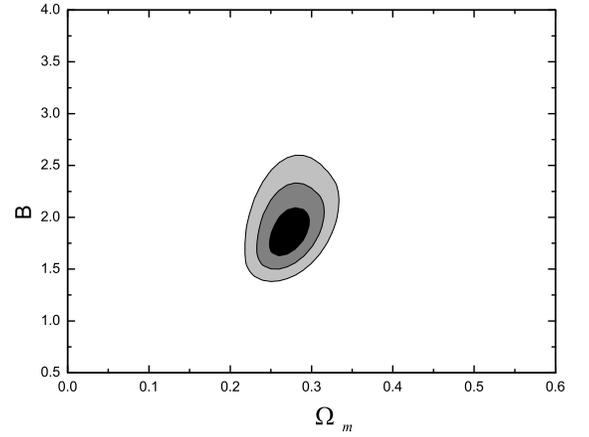}
\caption{The black, grey, and light grey region shows the 1, 2, and
3 $\sigma$ confidence level contours of $\Omega_m-B$ parameter space
respectively on combining the SNIa Gold data, CMB shift parameter,
and BAO.} \label{fig:Figure3}
\end{figure}

In Fig.3, we plot the results of the combined analysis of Riess SNIa
data + BAO + CMB. Again, the black, grey, and light grey region
shows the 1, 2, and 3 $\sigma$ confidence level contours on
$\Omega_m-B$ parameter space respectively with the minimum
$\chi^2=164.09$ occurring at \Om=0.27, and $B$=1.83 for EDE model.
We show clearly that combining two constraint 'bananas' from
SNIa/BAO+CMB can get tight constrains on both the matter content
\Om~ and parameter $B$ without significant degeneracy direction. On
the other hand, the two sets of constraints from SNIa/CMB+BAO are
largely consistent with each other, indicating the feasibility of
our $\Omega_m-B$ parametrization as a successful way to parameterize
our later universe. Considering the best fit values of \Om~ and
\Ol~, however, there exist some differences between the constraint
results from SNIa and CMB+BAO. Such discrepancies have also appeared
in data analysis on other cosmology models. These might imply the
existence of some systematics for cosmological observations we have
used here and/or potential inconsistencies which deserve further
investigations.

\begin{figure}[tbp]
\centering
\includegraphics[scale=0.8] {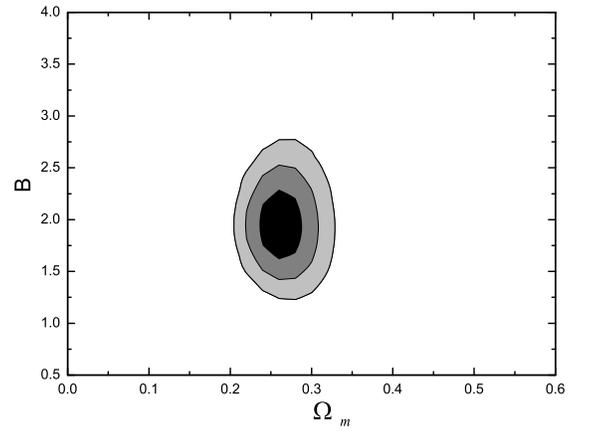}
\caption{The black, grey, and light grey region shows the 1, 2, and
3 $\sigma$ confidence level contours of $\Omega_m-B$ parameter space
respectively on combining the SNIa SNLS data, CMB shift parameter,
and BAO.} \label{fig:Figure4}
\end{figure}

Fig.4 shows the confidence contours of the combined analysis on EDE
model $\Omega_m-B$ parameter space combining CMB and BAO constraints
with SNLS data instead of Riess gold data. The minimum $\chi^2$
locates at \Om=0.26, and $B$=1.90. The constraints from combining
SNLS SNIa data with CMB+BAO are less restrictive than combining
Riess gold data, but more consistent with standard flat concordance
model. We note that the difference between two best fit parameter
values is due to difference between Riess gold data and SNLS data.
SNLS data gives the minimum $\chi^2=110.97$ occurring at \Om=0.30,
and $B$=2.35, much more consistent with concordance model than
result from Riess gold data (see figure 1).

In addition to cosmological constraints from kinetic distance
information obtained from different methods, it is especially
helpful to regard the structure formation process as a basis to test
our model by using e.g. gravitational lensing, galaxy cluster
abundance, galaxy clustering/dynamics and the CMB ISW effect.
Further tests are needed to discriminate our model from cosmology
models, such as DGP model. The CMB anisotropies and matter power
spectrum provide in principle suitable discriminatory tests. These
tests require a detailed understanding of the evolution of density
perturbations in our model. Fig.5 shows the linear growth factor
G(a) of $\Lambda$CDM model (red curve), DGP model (green curve) and
our model (black curve). The growth factor G(a) is defined by
solving the following differential equation\cite{para}

\beqa \frac{dG}{d\ln a}&+&\left(4+\frac{1}{2}\frac{d\ln H^2}{d\ln
a}\right)\,G
+G^2 \nonumber \\
&+&3+\frac{1}{2}\frac{d\ln H^2}{d\ln a}-\frac{3}{2}\om(a)=0,
\label{eq:main} \eeqa
where $G=d\ln(\delta/a)/d\ln a$, $H=\dot a/a$
is the Hubble parameter. The growth history for a flat universe can
be solved as
\beqa
G(a)=-1&+&[a^4H(a)]^{-1}\int_0^a \frac{da'}{a'}\,a'^4H(a') \nonumber \\
&\times&\left[\frac{5}{2}-\frac{3}{2}\ow(a')-G^2(a')\right].
\label{eq:gahw} \eeqa

For growth during the matter-dominated era, $G$ will be small. A
reasonable approximation throughout the growth history even as dark
energy comes to dominate has also been shown in \cite{para}

\beq G(a)=-\frac{1}{2}\ow(a)-\frac{1}{4}a^{-5/2}\int_0^a
\frac{da'}{a'}\,a'^{5/2} \ow(a'). \label{eq:gow} \eeq For any
particular model of $H(a)$, or $\om(a)$ or $\ow(a)$, we can then
evaluate the growth history.

The values of model parameters we chose to plot G(a) in Fig.5
correspond to the combined analysis results including CMB, BAO, and
Riess gold SNIa data. We can find that our best fit model mimic
$\Lambda$CDM linear structure formation quite well both in the early
universe and in the late universe. Just for comparison, we also show
the linear growth factor for DGP model. The non-linear structure
formation in our model is definitely worth to study but it is beyond
the scope of this present paper.

\begin{figure}[tbp]
\centering
\includegraphics[scale=0.6] {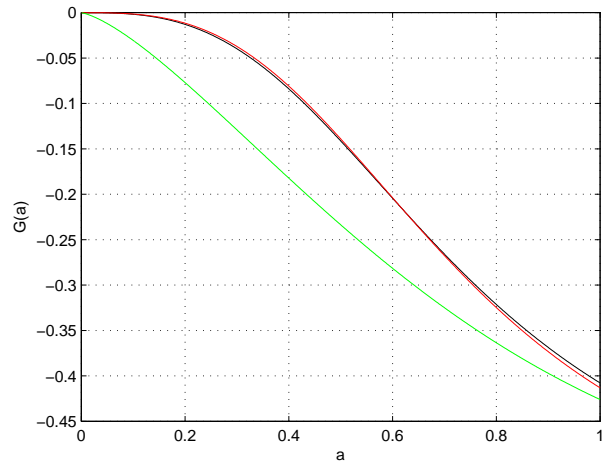}
\caption{The black, red, and green curve shows the linear growth
factor of our EDE model, $\Lambda\mathrm{CDM}$ and DGP model
respectively with best fitting parameter value from combined
analysis of SNIa Gold data, CMB shift parameter, and BAO.}
\label{fig:Figure5}
\end{figure}

In Fig.6, we plot the relative weight of EDE component and dark
matter component with respect to total energy contents in our
universe versus redshift with best fitting parameter value from
combined analysis of SNIa Gold data, CMB shift parameter, and BAO.
We can see that the DM-EDE equality time happened at z$\sim$0.7
which is quite close to the result from fitting to
$\Lambda\mathrm{CDM}$ model.

\begin{figure}[tbp]
\centering
\includegraphics[scale=0.55] {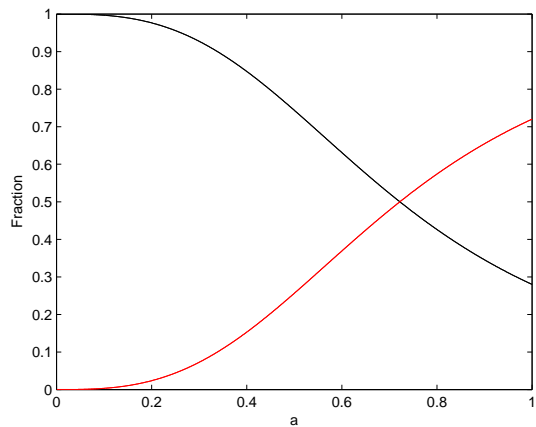}
\caption{The red and black curve shows the relative weight of EDE
component and dark matter component versus redshift respectively
with best fitting parameter value from combined analysis of SNIa
Gold data, CMB shift parameter, and BAO. The DM-EDE equality time
happened at z$\sim$0.7. } \label{fig:Figure6}
\end{figure}

\section{Parameter Degeneracy Analysis}

In this section, we discuss the new degeneracies on $\Omega_m-B$
parameter space, where new introduced parameter $B$ describes either
dark energy or curvature term plus the dark energy. In this section,
we use the first year of SNLS data in our analysis instead of Riess
gold sample, since SNLS data set has a relatively narrow redshift
range with z$\lesssim$1 thus with more clear degenerate features
between parameters. We note again that for the cosmological fits,
two of the SNLS data points were excluded because they are outliers
in the Hubble diagram. We also take the advantage of the recent GRB
sample compiled by Schaefer \cite{Schaefer} including 69 bursts with
properly estimated and corrected redshifts to investigate the
cosmological constraints. The redshift of the sample extends to
z$=$6.3 with considerable objects having z$>$1.5. Upon using these
distance modulus from GRBs, we fully aware of the circulation
problem associated with GRBs as cosmological probes. In this paper,
we only use the GRBs data to study the degeneracy properties on
$\Omega_m-B$ parameter space, but not combing to other cosmological
observations to constrain parameter space.

In our model, the parameter $B$ represents the deviation from
standard flat $\Lambda$CDM concordance model. we can easily find
that $B=2$ corresponding to $\Lambda$CDM model with cosmological
constant as dark energy and with flat geometry of our universe,
whereas $B>2$ describes effective positive curvature geometry of our
universe and/or effective dark energy equation of state $w<-1$,
namely phantom like dark energy, and $B<2$ describes effective
negative curvature geometry of our universe and/or effective dark
energy equation of state $w>-1$, namely quintessence like dark
energy. It is well known that there exists significant degeneracies
among $\Omega_k$, $\Omega_m$ and dark energy equation of state $w$
parameters. In the first step to explore evidences beyond standard
cosmology model, it might be helpful and reasonable to introduce
only one parameter which collapses both curvature effect and dark
energy effect into this single parameter, and maybe includes other
unknown features of new physics beyond flat $\Lambda$CDM concordance
model, simplify the degeneracy relations, thus make the signal of
deviation from flat $\Lambda$CDM model easily spotting out.

It is known that the CMB data alone cannot constrain well the
dynamics of dark energy. Additional information from large scale
structure of galaxies helps to tight on dark energy constraints
mostly because they provide a tight limit on $\Omega_m$, which in
turn helps to constrain the properties of dark energy due to
breaking the degeneracy between $\Omega_m$ and the equation of state
of dark energy in cosmological observable quantities. Here, we concentrate our
study on parameter degeneracies in luminosity/angular diameter
distance since it can be clearly and easily understood and it can also
give rise to the most direct constraints on dark energy models. With
flat universe assumption, the luminosity distance can be written as
\begin{equation}
d_L=c(1+z)\int^{z}_{0}\frac{d z^{\prime}}{H_0[\Omega_m
(1+z^{\prime})^3+(1-\Omega_m)(1+z^{\prime})^{2-B}]^{\frac{1}{2}}}.
\end{equation}
The degeneracies between $B$, $\Omega_m$ and $H_0$ are clearly seen
in this integral.

In order to see the degeneracy between the parameters $B$ and
$\Omega_m$, in Fig.7 we present the degeneracies in luminosity
distance on the $\Omega_m-B$ parameter plane at different redshifts.
The different color bands describe the parameter spaces of
$\Omega_m, B$ where given the variation of $d_L$ is in between $\pm
1 \%$ for $z=0.5$ (black), $0.1$(blue),$2$(yellow),
$3$(green),$6$(red),$1100$(magenta) with respect to a given fiducial
model with parameter value given by our best fittings from Riess
gold SNIa+CMB+BAO before. One can find that, the degeneracy between
$\Omega_m$ and $B$ varies with the redshift, which in turn implies
that combining the information of $d_L$ at different redshifts can
indeed helps broken such a degeneracy. This is the ideal case for
showing the degeneracy between the parameters $\Omega_m$ and $B$ for
different redshifts, because we fix the nuisance parameter $H_0$
instead of marginalizing it as we did in fitting procedure. Figure 8
is the results coming from the data fitting of GRBs (including high
redshift information up to $z\sim 6$) and SNLS with information from
much lower redshift range. We can find that the rotation of
degeneracy direction from low redshift to high redshift showing in
the plot can be explained by the degeneracy analysis on Fig. 7. The
trend of degeneracy rotation in $\Omega_m-B$ parameter plane is the
same for Fig.7 and 8. In order to constrain the cosmological
parameters $\Omega_m$ and $B$ well from only distance measurements,
one needs distance determinations for a wide range of redshifts. Or
instead, one can break the parameter degeneracies by other
cosmological observations with different degeneracy properties shown
in Fig.7. For current SNIa data, their redshift range is limited
with the highest observed redshift $\sim 1.7$ up to now. On the
other hand, for the GRB sample used in our analysis, the redshift
extends to as high as $\sim 6.3$. Due to the different degeneracies
at different redshift range, the complementarity of GRBs to SNIa is
highly expected with assumption of well controlled systematics of
using GRBs as standard candles.

We note that gravitational radiation opens another window by providing
high redshift information to constrain our model. Observations of
the gravitational waves emitted from the coalescence of supermassive
black holes with independent determination of redshift through an
electromagnetic counterpart can be used as standard sirens to
provide an excellent probe of the expansion history of the Universe,
especially by high redshift information, thus which can be used to
constrain the dark energy properties\cite{gw}. The degeneracy
properties of model parameters are the same as by using standard
candle, standard ruler or standard siren, as discussed in this
paper. Potentially, several well measured standard sirens will be
enough to give us tight constraints on dark energy parameters.

\begin{figure}[tbp]
\centering
\includegraphics[scale=0.43] {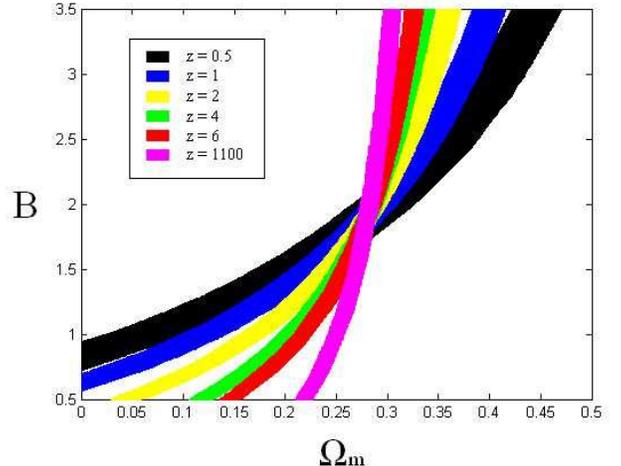}
\caption{The different color region simble $\pm 1\%$ variation
around lines of constant $d_L$ at redshift 0.1 (black),
0.5(red),1(green), 2(blue),3(cyan),10(magenta),1100(yellow), taking
fiducial model with best fitting combined analysis parameter value
from Gold sampe+CMB+BAO. This plot delineates the degeneracy between
the parameters $\Omega_m$ and $B$ at different redshift z.}
\label{fig:Figure7}
\end{figure}

In Fig. 9, we plot the effective dark energy equation of state
$w_{\text{eff}}$ with different choices of parameter B. The
effective dark energy equation of state is determined purely by the
Hubble paramter $H(z)$ , and there is a general formula that can
relate $H(z)$ and $w_{\text{eff}}$\cite{linjen} as
\begin{equation}
w_{\rm DE,eff}(z)\equiv -1+\frac{1}{3}\frac{d\ln(\delta H^2/H_0^2)}
{d\ln(1+z)}, \label{eq.wdh}
\end{equation}

\begin{figure}[tbp]
\centering
\includegraphics[scale=0.8] {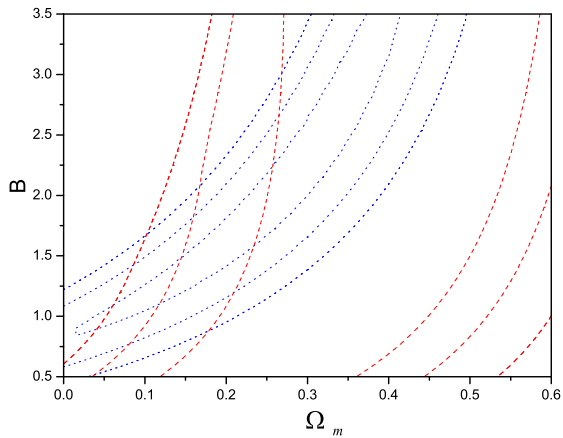}
\caption{1, 2, and 3 $\sigma$ confidence level contours on
$\Omega_m-B$ parameter space using the GRB sample and SNIa SNLS
data. The red-dashed lines are the results from the GRB sample and
the blue-dotted lines show the constraints resulting from the SNIa
SNLS data. } \label{fig:Figure8}
\end{figure}

\begin{figure}[tbp]
\centering
\includegraphics[scale=0.6] {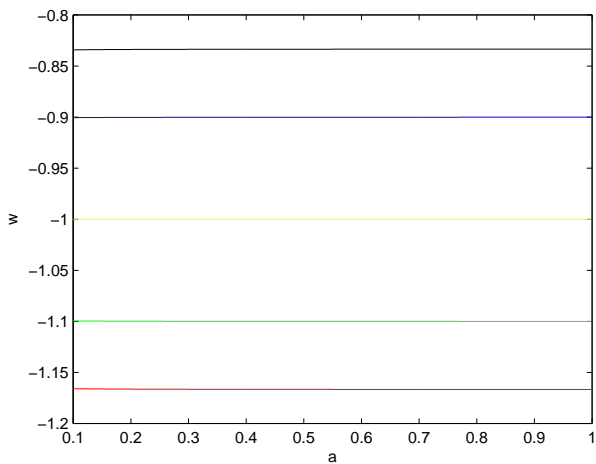}
\caption{Effective dark energy equation of state w(a). The different
color curves correspond to different value of B, namely $B=1.5$
(black), $B=1.7$(blue),$B=2.0$(yellow),
$B=2.3$(green),$B=2.5$(red).} \label{fig:Figure9}
\end{figure}

\section{Conclusion}

We have presented a cosmic model parameterizing the late universe
which collapses curvature and dark energy effects into one parameter
$B$ that may indicate any deviation from standard flat $\Lambda$CDM
model and we find that we can not conclude that the cosmic curvature
term is constantly zero, instead it may contribute rich
phenomenological effects. In order to show the advantages of our
parametrization, we study the degeneracy properties between $B$ and
$\Omega_m$, emphasizing the contribution from high redshift distance
information from GRB or gravitational waves experiments on-going and
up-coming. It is well-known that deducing the number of free
parameter without significant physics lost is quite important to
constrain cosmology models and to find new physics behind.

In this paper we also investigated the DGP cosmological model in the
simplest flat geometry case with the extra dimension contribution as
an effective "cosmological constant", compared with our
parameterized model and the reduction to the power-law $\Lambda$CDM
model for the 4D real Universe. We find that the DGP model even in
the simplest case is still an interesting candidate for the current
cosmic speed-up expansion mechanism at long distances, while we know
that in the short ranges the model behaves as 4D conventional
gravity. We will exploit the non-compact extra dimension to see its
possible existence signatures via cosmic effects in the general DGP
model later as a promising model, while we do not intent to discuss
the quantum aspects of this model as a basic theory\cite{alu}.

As a generalization of the $\Lambda$CDM model with naive
cosmological constant as dark energy candidate we has parameterized
a curvature like term with new phenomenological features via
numerical fittings and show the term explicit physics meanings when
we perform the parameter B reduction directly to zero or 2. It may
be interesting also to study the general properties of the
parameterized term as the matter-energy contents in our Universe
continuous equation to see what kind of "matter" it may describe effectively,
without specifying the form of the parameter. Besides, the
phantom case can be realized too, for example, the equation of state
parameter $w=p/\rho<-1$ if we take $B>2$ and quintessence
corresponds to $B<2$ with $w=p/\rho>-1$ numerically. We think this
picture is in conformity with other popular models and enlarges
phenomenological dark energy study possibilities to explain the
late-time accelerating expansion of our Universe, thus it is worth
of further endeavors.

\section*{Acknowledgements}
 X.-H.M. is
supported partly by NSFC under No. 10675062 and in part by the 2nd
stage Brain Korea 21 Program. M.S. acknowledges valuable discussions
with Zuhui Fan, Xuelei Chen, Pengjie Zhang, Hongsheng Zhao, Hong Li,
Tongjie Zhang and technical supports from Bo Liu. Besides, both
authors XHM and MS would like to thank AS-ICTP for the hospitality
where part of this work has been completed.


\begin{thebibliography}{99}
\bibitem{Perlmutter} S. Perlmutter el al. Nature 404 (2000) 955;
Astroph. J. 517 (1999) 565; A. Riess et al. Astroph. J. 116 (1998)
1009; astro-ph/0611572.

\bibitem{Spergel}
  D.~N.~Spergel {\it et al.},
  Astrophys.\ J.\ Suppl.\  {\bf 170}, 377 (2007).


\bibitem{page}
  L.~Page {\it et al.},
  Astrophys.\ J.\ Suppl.\  {\bf 170}, 335 (2007).


\bibitem{hinshaw}
  G.~Hinshaw {\it et al.},
  Astrophys.\ J.\ Suppl.\  {\bf 170}, 288 (2007).


\bibitem{jarosik}
  N.~Jarosik {\it et al.},
  Astrophys.\ J.\ Suppl.\  {\bf 170}, 263 (2007).


\bibitem{boomerang}
  C.~J.~MacTavish {\it et al.},
  Astrophys.\ J.\  {\bf 647}, 799 (2006).

\bibitem{cbi}
  A.~C.~S.~Readhead {\it et al.},
  Astrophys.\ J.\  {\bf 609}, 498 (2004).

\bibitem{sa}
  C.~Dickinson {\it et al.},
  Mon.\ Not.\ Roy.\ Astron.\ Soc.\  {\bf 353}, 732 (2004).

\bibitem{kuo}
  C.~l.~Kuo {\it et al.},
  Astrophys.\ J.\  {\bf 600}, 32 (2004).

\bibitem{planck}
  Planck Collaboration,
  arXiv:astro-ph/0604069;
J. Q. Xia, H. Li, G. B. Zhao, X. M. Zhang, arXiv:0708.1111; J. Bock
{\it et al.}, astro-ph/0604101;


\bibitem{cole}
  S.~Cole {\it et al.},
  Mon.\ Not.\ Roy.\ Astron.\ Soc.\  {\bf 362} (2005) 505.

\bibitem{sdss1} B. Roukema, et al., Astron. Astrophys. 382 (2002)
397

\bibitem{sdss2} D.J. Eisenstein et al., Astrophys. J. 633, 560
(2005);

\bibitem{sdss3}W. J. Percival, S. Cole, D. J. Eisenstein, R. C. Nichol, J. A.
Peacock, A. C. Pope, and A. S. Szalay, arXiv: 0705.3323; W. J.
Percival et. al., 2007, ApJ, 657, 51


\bibitem{tegmark}
  M.~Tegmark {\it et al.},
  Phys.\ Rev.\  D {\bf 74}, 123507 (2006).

\bibitem{Peebles} Linder, Dark energy resource letter, 0705.4102[astro-ph] and refereneces therein; S. Weinberg, Rev. Mod. Phys. {\bf 61}, 1 (1989);
P.J.E. Peebles and B. Ratra, astro-ph/0207347; S.M. Carroll, Living
Rev. Rel. 4 (2001) 1; T. Padmanabhan, Phys. Rep. 380 (2003) 235; T.
Padmanabhan, astro-ph/0603114; M. Kamionkowski, arXiv:0706.2986; S.
Nobbenhuis, Found. Phys. 36 (2006) 613; R. Bousso, arXiv:0708.4231;
J.-P. Uzan, arXiv:astro-ph/0605313; Albrecht, A. et al.,
astro-ph/0609591; J.Ren and X.H.Meng, Phys.Lett.B 633(2006)1 and
Phys.Lett. B636 (2006) 5; M.G.Hu and X.H.Meng, Phys.Lett.B
635(2006)186; X.H.Meng, J.Ren and M.G.Hu, Comm.Theor.Phys.
47(2007)379.

\bibitem{Lue} A. Lue, R. Scoccimarro and G. Starkman, astro-ph/0307034; X.H.Meng and P.Wang, Class. Quant.Grav. 20(2003)4949; ibid 21(2004)951;
ibid 21(2004)2029;ibid 22(2005)23; ibid, Phys. Lett. B 584(2004)1


\bibitem{Dvali} G. Dvali, G. Gabadadze and M. Porrati, Phys. Lett.
B 485 (2000) 208; D. Deffayet, Phys. Lett. B 502 (2001)199

\bibitem{rs} L. Randall, R. Sundrum, 1999, Phys. Rev. Lett. 83
(1999) 3370, hep-th/9905221; 83 (1999) 4690, hep-th/9906064.

\bibitem{hw} P. Ho\v rava, E. Witten, Nucl. Phys. B.
460 (1996) 506, hep-th/9510209; 475 (1996) 94, hep-th/9603142.

\bibitem{ms2} M.~Su, Z.~H.~Fan and Z. Wang in preparation

\bibitem{aa} A.Albrecht, 0710.0868[astro-ph]

\bibitem{distinguish} M. Ishak, A. Upadhye, D. N. Spergel, Phys. Rev. D.,74,
043513,(2006); M. Kunz, arXiv:astro-ph/0702615

\bibitem{curvature1}
G. B. Zhao, J. Q. Xia, H. Li, C. Tao, J. M. Virey, Z. H. Zhu and X.
Zhang, Phys. Lett. B 648, 8 (2007); K. Ichikawa, M. Kawasaki, T.
Sekiguchi, and T. Takahashi, JCAP 0612 (2006) 005

\bibitem{curvature}
C. Clarkson1, M. Cort and B. Bassett, astro-ph/0702670


\bibitem{we} P. Wang and X.H.Meng, Class.Quant. Grav. 22 (2005) 283; S. Ray, U.
Mukhopadhyay, X.H.Meng,  Gravitation and Cosmology, 13 (2007) 142;
M. Su, Z.-H. Fan, Z. Wang in preperation

\bibitem{dem} E.J. Copeland, M. Sami, S. Tsujikawa, hep-th/0603057 and references therein.


\bibitem{lensing3}
  K.~M.~Smith, W.~Hu and M.~Kaplinghat,
  Phys.\ Rev.\  D {\bf 74}, 123002 (2006).

\bibitem{lensing4}
  W.~Hu,
  Phys.\ Rev.\  D {\bf 65}, 023003 (2002).

\bibitem{lensing1}
  A.~Lewis and A.~Challinor,
  Phys.\ Rept.\  {\bf 429}, 1 (2006).


\bibitem{lensing2}
  M.~Zaldarriaga and U.~Seljak,
  Phys.\ Rev.\  D {\bf 58}, 023003 (1998).

\bibitem{lensing5}
W.~Hu, D.~Huterer, and K.~M. Smith, Astrophys.J. 650 (2006) L13

\bibitem{bispectrum}
M. Takada, \& B. Jain, 2004, MNRAS, 348, 897; E. Sefusatti, M.
Crocce, S. Pueblas, R. Scoccimarro, Phys. Rev. D 74, (2006) 023522

\bibitem{LF}
U. Seljak, A. Slosar, \& P. McDonald, JCAP 0610 (2006) 014

\bibitem{su3}

C. Alcock, \& B. Paczy\'{n}ski, 1979, Nature, 281, 358; L. Sun, M.
Su and Z. H. Fan, Chin. J. Astron. Astrophys. Vol. 6 (2006), No. 2,
155

\bibitem{WL}
D. Munshi, P. Valageas, L. Van Waerbeke, A. Heavens,
astro-ph/0612667; J. Benjamin et al., astro-ph/0703570; Jarvis M.,
Jain B., Bernstein G., Dolney D., 2006, ApJ, 644, 71; T. Kitching,
et al., 2006, MNRAS, 376, 771; H. Hoekstra et al., 2006, ApJ, 647,
116; C. Schimd et al., 2007, A\&A, 463, 405; S. Lee, K. W. Ng,
arXiv:0707.1730; W. Hu, \& B. Jain, Phys. Rev. D 70 (2004) 043009;
L. Amendola, M. Kunz, D. Sapone, arXiv:0704.2421;


\bibitem{su1}
  H.~Li, M.~Su, Z.~Fan, Z.~Dai and X.~Zhang,
  arXiv:astro-ph/0612060.

\bibitem{su2}
M. Su, H. Li, Z. H. Fan and B. Liu, astro-ph/0611155

\bibitem{Schaefer}
 B. E. Schaefer, Astrophys. J. 660, 16 (2007)

\bibitem{radiosources}
  J. C. Jackson, \& A. L. Jannetta, JCAP 0611 (2006) 002

\bibitem{fgas}
A. Mantz, S. W. Allen, H. Ebeling, D. Rapetti, arXiv:0709.4294

\bibitem{hubbleparameter}
L. Samushia, \& B. Ratra, Astrophys. J. 650 (2006) L5; H. Y. Wan, Z.
L. Yi, T. J. Zhang, J. Zhou, Phys. Lett. B 651 (2007) 352

\bibitem{clustercount}
W. Fang, \& Z. Haiman, Phys. Rev. D 75, 043010 (2007); M. Takada, \&
S. Bridle, arXiv:0705.0163; W. Hu, Phys. Rev. D 67, 081304 (2003);
M. Lima, \& W. Hu, Phys. Rev. D 70, 043504 (2004)


\bibitem{Riess}
  A.~G.~Riess {\it et al.},
  Astrophys.\ J.\  {\bf 659}, 98 (2007)

\bibitem{SNLS}
P. Astier, J. Guy, N. Regnault et al., 2006, A\&A, 447, 31

\bibitem{R} Y. Wang \& P. Mukherjee, 2006, ApJ, in press, astro-ph/0604051


\bibitem{Hubble}
  W.~L.~Freedman {\it et al.},
  Astrophys.\ J.\  {\bf 553}, 47 (2001)

\bibitem{para}
E. V. Linder, \& R. N. Cahn, arXiv:astro-ph/0701317


\bibitem{MG}
H. Jain, \& P. J. Zhang, arXiv:0709.2375; S. Wang, L. Hui, M. May,
Z. Haiman, Phys. Rev. D76 (2007) 063503; W. Hu, I.Sawicki,
arXiv:0708.1190; E. Bertschinger, Astrophys. J. 648 (2006) 797;  P.
J. Zhang, M. Liguori, R. Bean, S. Dodelson, arXiv:0704.1932; Y.-S.
Song, W. Hu, I. Sawicki, Phys. Rev. D 75 (2007) 044004; D. Huterer,
\& E. V. Linder, Phys. Rev. D75 (2007) 023519; M. Ishak, A. Upadhye,
D. N. Spergel, Phys. Rev. D74 (2006) 043513; L. Knox, Y.-S. Song, J.
A. Tyson, arXiv:astro-ph/0503644; C. Skordis, Phys. Rev. D74 (2006)
103513;  V. Sahni, Y. Shtanov, A. Viznyuk, JCAP 0512 (2005) 005; M.
White, \& C. S. Kochanek, Astrophys.J. 560 (2001) 539; A. Shirata,
T. Shiromizu, N. Yoshida, Y. Suto, Phys. Rev. D71 (2005) 064030; C.
Sealfon, L. Verde, R. Jimenez, Phys. Rev. D71 (2005) 083004; H. F.
Stabenau, \& B. Jain, Phys. Rev. D74 (2006) 084007; A. Shirata, Y.
Suto, C. Hikage, T. Shiromizu, N. Yoshida, arXiv:0705.1311; M. Kunz
\& D. Sapone, Phys. Rev. Lett. 98 (2007) 121301;


\bibitem{gw} B. Kocsis, Z. Frei, Z. Haiman and K. Menou, ApJ.
637 (2006) 27; N. Dalal, D. E. Holz, S. A. Hughes and B. Jain, Phys.
Rev. D., 74 (2006) 063006; C. Deffayet, K. Menou, arXiv:0709.0003;


\bibitem{linjen}
E. V.\ Linder \& A.\ Jenkins, MNRAS 346, 573 (2003)



\bibitem{smartidea}S. Sullivan,
A. Cooray, D. E. Holz, arXiv:0706.3730 [astro-ph]; Narrowing
Constraints with Type Ia Supernovae: Converging on a Cosmological
Constant, JCAP09(2007)004

\bibitem{PM}
 R. de Putter \& E. V. Linder, arXiv:0710.0373; S. Sullivan, D. Sarkar, S. Joudaki,
A. Amblard, D. Holz, A. Cooray, arXiv:0709.1150;

\bibitem{alu}
A. Lue, Phys. Rept. 423,1 (2006)

\end{thebibliography}
\end{document}